\documentclass{cpcauth}
\usepackage{epsf,psfig,epsfig}
\usepackage{amssymb}

\newcommand{\be}{\begin{equation}}
\newcommand{\ee}{\end{equation}}
\newcommand{\bea}{\begin{eqnarray}}
\newcommand{\eea}{\end{eqnarray}}

\begin{document}

\begin{frontmatter}

\title{Go with the Winners: a General Monte Carlo Strategy} 

\author{Peter Grassberger}

\address{John von Neumann - Institut f\"ur Computing,\\
     Forschungszentrum J\"ulich, D-52825 Germany}

\begin{abstract}
We describe a general strategy for sampling configurations from a given
distribution, {\it not} based on the standard
Metropolis (Markov chain) strategy. It uses the fact that nontrivial 
problems in statistical physics are high dimensional and often close 
to Markovian. Therefore, configurations are built up in many, usually 
biased, steps. Due to the bias, each configuration carries its
weight which changes at every step. If the bias is close to optimal, all 
weights are similar  and importance sampling is perfect. If not,
``population control" is applied by cloning/killing partial configurations 
with too high/low weight. This is done such that the final (weighted) 
distribution is unbiased. We apply this method (which is also closely 
related to diffusion type quantum Monte Carlo) to several problems of 
polymer statistics, reaction-diffusion models, sequence alignment, 
and percolation.

\end{abstract}

\end{frontmatter}

\section{Introduction}

Although Markov chain (Metropolis-type) Monte Carlo (MC) simulations dominate
in statistical physics today, simulations not based on this strategy have
been used from early times on. Well known examples are evolutionary (in particular
genetic) algorithms \cite{schwefel}, diffusion type quantum 
MC \cite{vonderlinden}, and several 
algorithms devised for the simulation of long chain molecules 
\cite{rosenbluth,wall,redner-rey,garel,perm}.

As these methods were developed 
independently in different communities, it was not generally recognized
-- or rather forgotten -- that most of them are realizations of a common 
strategy, as pointed out by 
Aldous and Vazirani \cite{aldous} who also coined the name ``go with
the winners". But essentially the same basic strategy was already discussed as 
a general purpose sampling method by Herman Kahn in 1956 \cite{kahn} who 
called it ``Russian Roulette and Splitting", and attributed it to unpublished 
work by von Neumann and Ulam. For further applications of this 
strategy see \cite{perm-gfn,iba,liu}. The last two references also discuss 
applications in lattice spin systems and Bayesian inference, fields which will 
not be treated in the present review.

\section{The Basic Strategy}

\subsection{Sequential Importance Sampling}
\vspace*{-3mm}
As in any MC method, we draw configurations ${\bf x}$ from some distribution $p({\bf x})$.
Writing the partition sum as 
\be
   Z = \sum_{\bf x} e^{-\beta H({\bf x})} \approx M^{-1} \sum_{\alpha = 1}^M
         e^{-\beta E({\bf x}^\alpha)}/p({\bf x}^\alpha) ,
\ee
we can interpret this as each configuration having its {\it weight} 
$W({\bf x}) = e^{-\beta E({\bf x})}/p({\bf x})$. In importance sampling we try to 
choose $p({\bf x}) \propto e^{-\beta E({\bf x})}$, so that all weights become 
equal. 

We now assume that we can break up the construction of a configuration into 
$N$ single steps $x_n, n=1, 2, \ldots N$. For a polymer, the $n$-th step would
e.g. be the placement of the $n$-th monomer. Then the weight $W$ is obtained 
recursively as $W=W_N$, $W_0=1$,
\be
   W_n = W_{n-1} {e^{-\beta (E(x_1, \ldots x_n)-E(x_1, \ldots x_{n-1}))} \over
     p_n(x_n|x_1 \ldots x_{n-1})}\;.
\ee
In statistics this is called sequential importance sampling (SIS) \cite{liu}. 

In some cases, ``natural" values for the $p_n(x_n)$ are easy to guess. In the 
{\it Rosenbluth} method \cite{rosenbluth} for simulating a self-avoiding walk (SAW),
e.g., one chooses uniformly among the free neighbours. But this is not optimal, a
better choice is provided by {\it Markovian anticipation} \cite{cylinder}.
In general, for choosing $p_n(x_n|x_1 \ldots x_{n-1})$ one has to depend on heuristics,
except in the case of diffusion quantum MC where perfect importance 
sampling ($W_N\equiv 1$) is possible if the ground state wave function is 
known \cite{vonderlinden,chemnitz}.
Specific choices will be discussed together with applications.

\subsection{Population Control}
\vspace*{-3mm}
The main drawback of SIS is that the distribution
of weights can become extremely wide. If long range correlations are weak
(as e.g. for SAWs), $\log W_N$ is roughly a sum of independent terms. 
This suggests the following strategy:

If at any step $n$ the weight $W_n$ is above a suitably chosen threshold 
$W_n^+$, we make an additional copy of the configuration $x_1, \ldots x_n$, 
and give both copies the weight $W_n/2$. Both are then
grown independently (with eventual later copyings) up to full length 
\footnote{In some cases (e.g. at low temperatures, where Boltzmann factors 
are huge), it might be necessary to make several copies and to distribute
the weight evenly among them.}.
In this way high weights are suppressed and precious ``good" configurations 
are less likely to be lost entirely by bad subsequent moves. In \cite{wall} a 
similar strategy (but not based on weights) was called `enrichment'.

On the other hand, if $W_n$ falls below another threshold $W_n^-$, we draw 
a random number $r \in [0,1]$. If $r<1/2$ we kill the configuration and start 
a new one. If $r>1/2$ we keep it and double its weight.

Obviously, for any choice of the thresholds, neither the cloning nor the 
pruning introduce any additional bias. Thus 
we can, in principle, use any choice for $W_n^+$ and $W_n^-$, and we can 
change them ad libitum during the simulation. Bad choices will, however, 
lead to inefficiency, just as do bad choices for $p_n(x)$.

Except at very low temperatures where special care is needed \cite{stiff,fold}
we found the following strategy to be sufficient:
\begin{itemize}
\item For the first configuration(s) we do not clone at all and kill only if 
the weight is exactly zero.
\item If we have already $m$ previous configurations which had reached 
size $\ge n$, we estimate from them the partition sum $Z_n \approx \hat{Z}_n 
\equiv  m^{-1} \sum_{\alpha = 1}^m W_n({\bf x}^\alpha)$. We then set
$W_n^\pm = C^\pm \hat{Z}_n$ with $C^+\approx 1/C^- \approx O(1) - O(10)$.
\end{itemize}

\subsection{Depth First Versus Breadth First}
\vspace*{-3mm}
As described above, the algorithm is most efficiently implemented in a 
{\it depth first} fashion, and as such was called PERM (pruned-enriched
Rosenbluth method) in \cite{perm}.
In a depth first approach \cite{tarjan}, we follow one copy until its 
end before we take up the other copy. In {\it breadth first} search, on 
the other hand, we treat all copies in parallel and handle the 
$n$-th steps of all copies before we go to $n+1$.

Evolutionary algorithms \cite{schwefel} are usually implemented breadth
first. One puts up a population of $M$ replicas
which are evolved simultaneously, and population control is exercized such 
that $M$ stays constant during the evolution. The same is true for most 
implementations of the ``go with the winners" strategy. This has several 
advantages:
\begin{itemize}
\item Breadth first approaches are well adapted for massively parallel 
computers. One simply puts one configuration on each processor.
\item One has no problem with keeping the number of replicas constant.
\item One can use more general population control strategies \cite{liu}.
\end{itemize}
But the last two points seem minor in most applications we have studied.
On the other hand, the main advantage of depth first is the elegance and 
efficiency of the codes. The easiest implementation is by means of 
recursion (for a pseudocode see \cite{perm}). Copies (or rather instructions
to make copies) are then put on a stack which is maintained automatically 
if recursive function calls are used. Storage use is minimized
(as only a single copy and its history is kept in memory), 
and communications are also less than in breadth first.

\section{Multiple Spanning Percolation Clusters}

Let us consider percolation on a large but finite rectangular
lattice in $2\leq d < 6$. We single out one direction as ``spanning". In this 
direction boundaries are open, while periodic b.c. are used in the other
direction(s). For a long time it was believed that there is at most one 
spanning cluster (which touches both open boundaries) 
in the limit of large lattices,
keeping the aspect ratio fixed ($L_i = x_i L,\; L\to\infty,\; i=1,
\ldots d$). 

\begin{figure}[b]
\begin{center}
\epsfig{figure=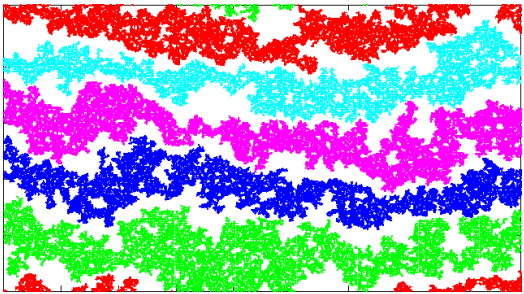,angle=270,width=.43\textwidth}
\end{center}
\caption[]{Configuration of 5 spanning site percolation clusters on a
    lattice of size $500\times 900$. Any two clusters keep a distance of at
    least 2 lattice units. Lateral boundary conditions are periodic.
    The probability to find 5 such spanning clusters
    in a random disorder configuration is $\approx 10^{-92}$.}
\label{figs5}
\end{figure}

Since there is no spanning cluster for subcritical percolation and 
exactly one in supercritical, the only relevant case is critical
percolation. There it is now known that the probabilities $P_k$ to have
exactly $k$ spanning clusters are all non-zero in the limit $L\to\infty$. In
$d=2$ they are known exactly from conformal invariance, but
for $d\geq 3$ no exact results are known. But there 
is a conjecture by Aizenman \cite{aizenman}, stating that for a lattice
of size $L\times \ldots \times L \times (rL)$ ($rL$ is the length in the
spanning direction) $   P_k \sim e^{-\alpha r}  $    with
\be
   \alpha \propto k^{d/(d-1)} \qquad {\rm for} \;\; k\gg 1.   \label{aizen}
\ee

For $d=2$ one has $\alpha \sim k^2$, in agreement with Eq.~(\ref{aizen}).
A generalization in $d=2$ consists in demanding that clusters are 
separated by at least $q$ paths on the dual lattice \cite{aizen-dupl-ahar}.
In that case, and for periodic transverse b.c.,  
\be
   \alpha = {2\pi\over 12}[((q+1)k)^2-1] \;.
         \qquad k\geq 2, d=2                                     \label{ada}
\ee
In order to test Eqs.~(\ref{aizen}),(\ref{ada}) for a wide range of values of
$k$ and $r$, one has to simulate events with tiny probabilities,
$\ln P_k \sim -10^2$ to $ -10^3$. It is thus not surprising that previous
numerical studies have verified Eq.~(\ref{ada}) only for small values of
$k$, and have been unable to verify or disprove Eq.~(\ref{aizen})
\cite{sen,shchur}.

To demonstrate that such rare events can be simulated with PERM,
we show in Fig.~2 a lattice
of size $500\times 900$ with 5 spanning clusters which keep distances $\geq 2$.
Eq.~(\ref{ada}) predicts for it $P_k = \exp(-336\pi/5) \approx 10^{-92}$.
This configuration was obtained by letting 5 clusters grow simultaneously,
using a standard cluster growth algorithm \cite{leath}, from the left border.
Precautions were taken that they grew with the same speed towards the right,
i.e. if one of them lagged behind, the growth of the others was stopped until
the lagging cluster had caught up. If one of them died, or if two came closer
than two lattice units, the entire configuration was discarded. If not, it was 
cloned if the weight $W_n$ exceeded $3\hat{Z_n}$. Note that here the growth
was made without bias, and therefore no pruning was necessary.

In this way we could check Eq.~(\ref{ada}) with high precision, proving 
the correctness of our algorithm.

More interesting is the test of Eq.~(\ref{aizen}) for $d=3$. Simulating up 
to 16 parallel clusters on lattices of sizes up to $128\times 128\times 2000$ 
(leading to probabilities as small as $10^{-300}$!) gave perfect agreement
with Eq.~(\ref{aizen}) \cite{gra-zif}.

\section{Polymers}

One of the main applications of the go-with-the-winners strategy is
configurational statistics of long polymer chains. For a breadth first 
algorithm which otherwise is very similar to PERM see \cite{garel}.

\subsection{$\Theta$-Polymers}
\vspace*{-3mm}
PERM is particularly efficient near the so-called `theta-' or coil-globule
transition. According to the generally accepted scenario, the
theta-point is tricritical with upper critical dimension $d_c=3$
\cite{degennes}.

At $T_\theta$, bias correction and Boltzmann factors nearly cancel in $d=3$.
Therefore, polymers have essentially random walk configurations with
small (logarithmic) corrections. Therefore, a non-reversing random walk
(U-turns are forbidden) for SIS is
already sufficient to give good statistics with very few pruning and
enrichment events. In \cite{perm} chains made of up to 1,000,000 steps
could be sampled with high statistics within modest CPU time \cite{perm}.
They were done in finite volumes (``dense limit") and verified
that the $\Theta$-point indeed is a second order transition.
The most precise verification of logarithmic 
corrections came from chains with $N=10,000$ in infinite volume. The 
deviations from random walk behaviour turned out to be much stronger than 
the leading-log corrections predicted from the renormalization group 
\cite{duplant-unmix}, but agreement improves substantially
when higher order corrections are included in the latter \cite{hager}.

\subsection{Critical Unmixing}
\vspace*{-3mm}
A related problem is the unmixing of semidilute polymer solutions. For 
any finite chain length $N$ this is in the Ising universality class. But
in addition to the Ising scaling laws, there are further universal 
scaling laws for parameters and amplitudes which, from the Ising
point of view, would be non-universal. In particular, the critical 
temperature should approach $T_\theta$ when $N\to\infty$,
$  T_c - T_\theta \sim N^{-1/2}  $, 
and the critical monomer concentration should tend to zero,
\be
   \phi_c \sim N^{-1/2}.                      \label{mix-phi}
\ee
The exponents here are mean field, appropriate for $d=3$. Indeed one 
should also expect logarithmic corrections \cite{duplant-unmix}.
Previous experiments had suggested an exponent $0.38\pm 0.01$ in 
Eq.~(\ref{mix-phi}). This 
would be very hard to understand and has stirred a lot of theoretical 
activity (for a review see \cite{widom,mix}). Simulations using PERM 
\cite{mix} showed that this is wrong: The deviations from Eq.~(\ref{mix-phi})
can be understood most easily as logarithmic corrections.

\subsection{DNA Melting}
\vspace*{-3mm}
DNA in physiological conditions forms a double helix. Changing the pH value
or increasing $T$ can break the hydrogen bonds between the base
pairs, and a phase transition to an open coil occurs. Experiments suggest
it to be first order \cite{wartell}. While a second
order transition would be easy to explain \cite{poland,fisher},
no previous model had been able to give a first order transition.

\begin{figure}[b]
\begin{center}
\epsfig{figure=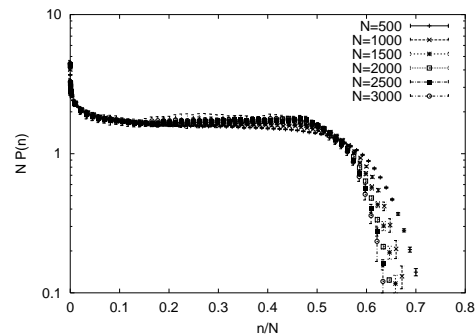,angle=270,width=.4\textwidth}
\end{center}
\caption[]{Histograms of the number of contacts, for single strand 
    length $N=500, \ldots 3000$, at $\epsilon=\epsilon_c$.
    On the horizontal axis is plotted $n/N$ as is appropriate for
    a first order transition.}
\label{figs910}
\end{figure}

The model studied in \cite{dna} lives on a simple cubic lattice. A double
strand of DNA with length $N$ is described by a diblock copolymer of
length $2N$, made of $N$ monomers of type $A$ and $N$ monomers of type $B$.
All monomers have excluded volume interactions, i.e. two monomers cannot
occupy the same lattice site, with one exception: The $k$-th $A$-monomer
and the $k$-th $B$-monomer, with $k$ being counted
from the center where both strands are joined together, can occupy the same
site. If they do so, then they even gain an energy $-\epsilon$. This models the
binding of complementary bases.

The surprising result of simulations of chains with $N$ up to 4000 is that
the transition is first order, but shows finite scaling behaviour as expected
for a second order transition with cross-over exponent $\phi=1$. To demonstrate
this, we show in Fig.~4 energy histograms for different chain lengths. One sees 
two maxima, one at $n=0$ and the other at $n\approx N/2$, whose distance scales
proportionally to $N$. But in
contrast to usual first order transitions the minimum in between 
does not deepen with increasing $N$. This is due to the absence of any
analogon to a surface tension. The same conclusion is obtained from specific
heat data and from end-to-end distances \cite{dna}. 

In \cite{dna} we also studied similar models with (partially) switched off
excluded volume effects. They show that excluded volume is the 
main force making the transition first order, as also confirmed by 
subsequent analytic calculations \cite{kafri-muk-peliti}.

\subsection{Native Configurations of Toy Proteins}
\vspace*{-3mm}
Predicting the native ($\approx$ ground) states of proteins is one of the most 
challenging problems in mathematical biology \cite{creighton}. It is difficult 
because of the many local energy minima. 

In view of this, there exists a large literature on finding
ground states of artificially constructed heteropolymers. Most of these
models are formulated on a (square or simple cubic) lattice and use only
few monomer types. The best known example is the HP model of K. Dill
\cite{dill85} which has two types of amino acids: hyrophobic (H) and
hyrophilic (polar, P) ones. With most algorithms, one can find ground states
typically for random chains of lengths up to $\sim 50$.

In \cite{fold} we used PERM to study several sequences, of the HP model
and of similar models, which had been discussed previously by other authors.
In all cases we found the known lowest energy states,
but in several cases we found new ones. A particularly impressive example 
is a chain of length 80 with two types of monomers, 
constructed such that it should fold into a bundle of four ``helices" with
an energy $-94$ \cite{otoole}. Even with a specially designed algorithm,
the authors of \cite{otoole} were not able to recover this state. With PERM
we not only found it easily, we also found several lower states, the lowest
one having energy $-98$ and a completely different structure.

\subsection{Miscellaneous}
\vspace*{-3mm}
Applications of PERM to other polymer problems are treated in
\cite{stiff,parallel,cylinder,manhattan,localize,hetero,prellberg}.
For problems with open coils, a bias strategy called {\it Moarkovian anticipation}
in \cite{cylinder} worked very well. Integrating over the disorder, we recently
could also map a biased random walker in the presence of random traps onto
a stretched collapsed polymer \cite{vishal}. Without bias, the transition 
from the finite time {\it Rosenstock} to the asymptotic {\it Donsker-Varadhan}
behaviour is in 3d akin to a cross-over in a first order phase transition. With
bias, the delocalization (globule-stretch) transition is first order in $d\ge 2$
\cite{vishal}.

\section{Lattice Animals (Randomly Branched Polymers)}

Consider the set of all connected clusters of $n$ sites on a regular lattice,
with the origin being one of these sites, and with a weight defined on
each cluster. Lattice animals are defined by giving the same
weight to each cluster. This distinguishes them from percolation clusters 
where the weight depends on the `wetting' probability $p$.
In the limit $p\to 0$ this difference disappears, and the two statistics 
coincide. It is believed that lattice animals are a good model for randomly 
branched polymers \cite{lubensky}.
While there existed no efficient algorithm for estimating the animal partition 
sum there exist very simple and efficient Leath-type \cite{leath} algorithms 
for percolation clusters.

\begin{figure}
\begin{center}
\parbox{.45\textwidth}{\epsfig{figure=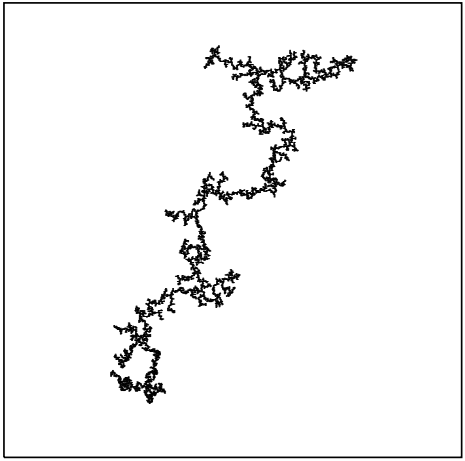,angle=270,width=.42\textwidth}}
\parbox[t]{.42\textwidth}{\caption[]{A typical lattice animal with 8000 sites
on the square lattice.}}
\end{center}
\label{fig14}
\end{figure}

Our PERM strategy \cite{chemnitz,animals} consists in starting off to generate 
subcritical percolation clusters by a (breadth first) Leath method, re-weighing 
them as animals while they are still growing, and in making clones of ``good"
ones. Since we work at $p<p_c$, we do not need pruning. The threshold $W_+$ for
cloning is chosen such that it depends both on the present animal weight and on 
the anticipated success for further growth.

In this way we obtained good statistics for animals of several thousand
sites, independent of the dimension of the lattice. A typical 2-$d$ animal with
8000 sites is shown in Fig.~5. We also simulated animal collapse (when
each nearest neighbor pair contributes $-\epsilon$ to the energy), and animals
near an adsorbing surface \cite{animals}.

\section{Error Estimates and Reliability Tests}

Statistical errors can be estimated as usual by dividing a long run into several
bunches, computing averages over each bunch, and studying the fluctuations
between them. For PERM the situation is indeed rather easy, since each {\it tour}
(set of all configurations generated by cloning from one common ancester)
is independent of any other.

To check for excessive fluctuations in weights $W$ of entire tours, we make 
a histogram on a logarithmic
scale, $P(\log(W))$, and compare it with the weighted histogram $WP(\log(W))$.
If the latter has its maximum for values of $\log(W)$ where the former is
already large (i.e. where the sampling is already sufficient), we are
presumably on the safe side. However, if $WP(\log(W))$ has its maximum
at or near the upper end of the  sampled range, we should be skeptical.

\begin{figure}
\begin{center}
\parbox{.23\textwidth}{\includegraphics[width=.20\textwidth,angle=270]{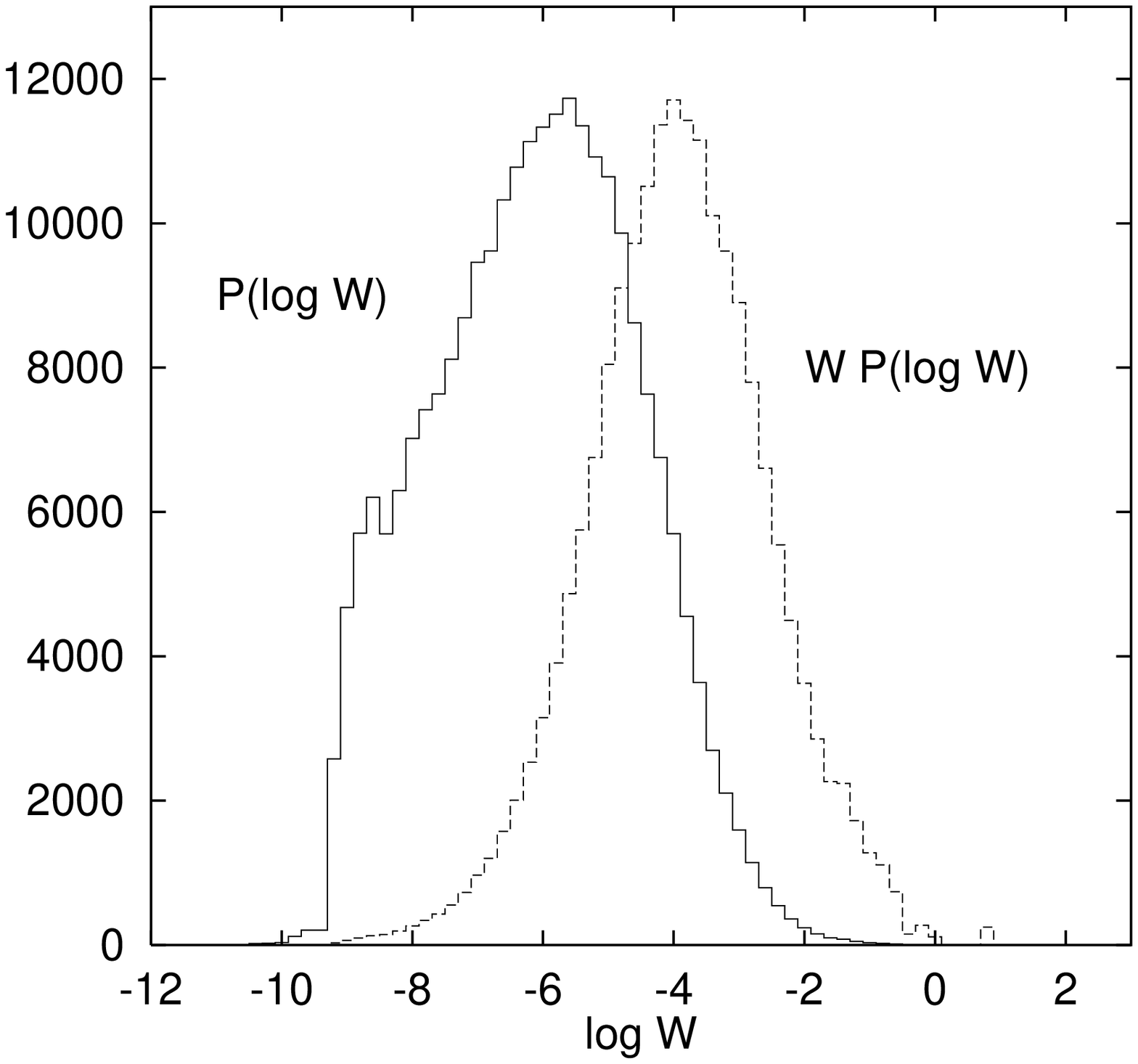}}
\parbox{.23\textwidth}{\includegraphics[width=.20\textwidth,angle=270]{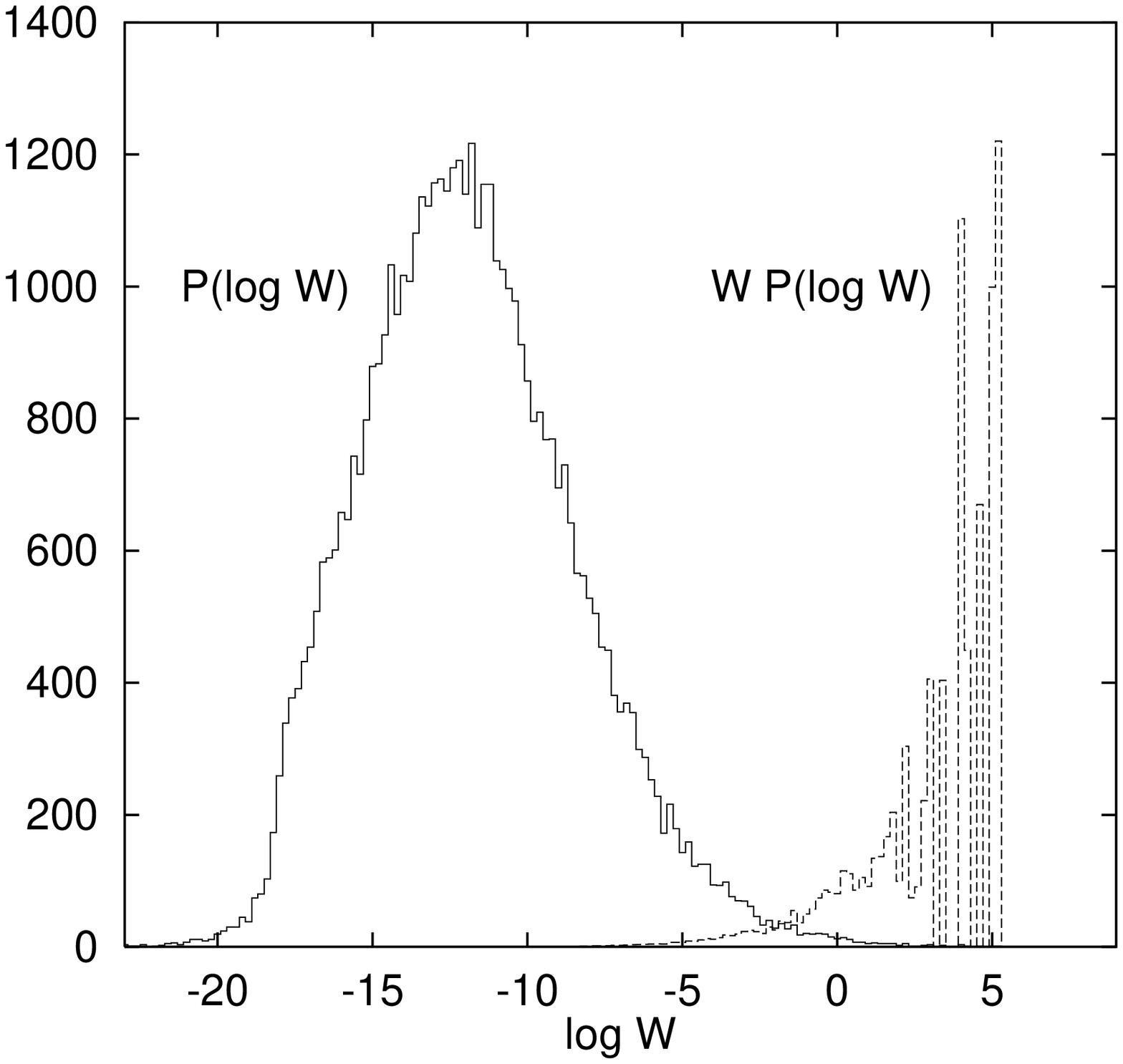}}
\caption[]{Full lines are histograms of logarithms of tour weights, normalized as
tours per bin.  Broken lines show the corresponding weighted distributions,
normalized so as to have the same maximal heights. Weights $W$ are only
fixed up to a $\beta$-dependent multiplicative constant. While the left panel 
suggests a reliable simulation, the right one was indeed wrong (from ref. 
\cite{localize}).}
\end{center}
\label{fig1617}
\end{figure}

In Fig.~6 we illustrate this with two figures taken from \cite{localize}.
While the left panel gave rise to correct results, the right one did not.

\section{Conclusion}

We have seen that MC simulations {\it not} following the 
Metropolis scheme can be very efficient. We have illustrated this with a
wide range of problems. Conspicuously, the Ising model was not among them.
It simply would be very hard to beat, say, the Swendsen-Wang
algorithm. In principle, the go-with-the-winners strategy has as wide
a range of applications as the Metropolis scheme. Its only requirement
is that instances (configurations, histories, ...) are built up in small steps,
and that the growth of their weights during the early steps of this build-up is
not too misleading.

The method is not new. It has its roots in algorithms which have been regularly 
used for several decades. Some of them, like genetic algorithms, are familiar to
most scientists, but it is in general not well appreciated that they can be
made into a general purpose tool. And it seems even less appreciated how
closely related are methods developed for quantum MC simulations, polymer 
simulations, and optimization methods. I firmly believe that this close relationship
can be made use of in many more applications to come.

Among these are significance tests for sequence alignment, where one needs 
large samples of 
random pairs of seqences in order to check whether an observed alignment is 
significant. Instead of really drawing random pairs, one can use PERM to 
draw biased pairs which are more similar than random ones, enhancing thereby
the interesting high-score region \cite{bundschuh-gr}.

Another application is to epidemic models where one can follow the fate 
of epidemics which have a very low chance of survival since, e.g., they 
started in a very hostile environment which they first have to adopt to their
needs. Here simulations with PERM \cite{unpubl} allowed to verify with 
very high statistics the claim of \cite{gcr} that no power laws result,
in contrast to previous suggestions. A final application to a toy `population'
model \cite{redner-krip} is discussed in \cite{chemnitz}.

\end{document}